\documentclass[%
 reprint,
 amsmath,amssymb,
 aps,
]{revtex4-2}

\usepackage[dvipdfmx]{graphicx}% Include figure files
\usepackage{dcolumn}% Align table columns on decimal point
\usepackage{bm}% bold math
\usepackage{color}
\begin{document}

\preprint{APS/123-QED}

\title{
Large-scale spatio-temporal patterns in a ring of non-locally
coupled oscillators with a repulsive coupling} 
%Analogy between multi-twisted states in a coupled oscillator array\\and directed percolation with two absorbing states}% Force line breaks with \\

\author{Bojun Li}
\author{Nariya Uchida}
\email{uchida@cmpt.phys.tohoku.ac.jp}
\affiliation{Department of Physics, Tohoku University, Sendai 980-8578, Japan}

\date{\today}% It is always \today, today,

\begin{abstract}
Non-locally coupled oscillators with a phase lag exhibit various non-trivial spatio-temporal patterns such as the chimera states and the multi-twisted states. 
We numerically study large-scale spatio-temporal patterns in a ring of oscillators with a 
repulsive coupling with a phase delay parameter $\alpha$.
We find that the multi-chimera state disappears when $\alpha$ exceeds a critical value. 
Analyzing the density of incoherent regions, we show that the transition is analogous to that of directed percolation with two absorbing states, but their critical behaviors are different.
The multi-chimera state reappears when $\alpha$ is further increased, exhibiting non-trivial spatio-temporal patterns with a plateau in the density of incoherent regions. A transition from the multi-chimera to multi-twisted states follows at a larger value  of $\alpha$, resulting in five collective phases in total.
\end{abstract}

\maketitle

%\tableofcontents

\section{Introduction}
The collective behavior of coupled oscillators is ubiquitous in Nature \cite{book1,book2}. 
When the coupling is weak, the dynamics of oscillators are described only by a phase parameter by means of phase reduction. 
Depending on the phase lag and range of coupling, the coupled phase oscillators exhibit not only global synchronization
and desynchronization~\cite{Acebron2005} 
but also a variety of patterns, such as 
the chimera states~\cite{kuramoto2002,Abrams2004,Bera2017,Omelchenko2018,suda2018breathing} and 
the twisted states~\cite{Wiley2006,Girnyk2012,Omelchenko2014,Laing2016}.
The chimera states are characterized by coexistence of coherent and incoherent regions,
while the twisted state is a state in which the phase difference between two neighbor oscillators is constant.
A widely used model to study these two states are 
the non-locally coupled phase oscillators on a one-dimensional ring, 
the time evolution of which obeys 
\begin{equation}
\label{eq.main}
\dot{\psi}(x,t)=\omega_0-\frac{1}{2R}\sum_{\substack{y=x-R\\y\ne x}}^{x+R}\sin\left[\psi(x,t)-\psi(y,t)+\alpha\pi\right],
\end{equation}
where $\psi(x,t)$ represents the phase of the oscillator at $x=1,2,\ldots, N$ at time $t$, $R$ is the coupling range, and $\alpha \pi$ gives the phase lag. 
The intrinsic phase velocity $\omega_0$ is constant and set to zero without losing generality by the transformation $\psi \to \psi - \omega_0 t$. 
    
The collective behavior of the oscillator system governed by Eq.~(\ref{eq.main}) 
is summarized as follows.
%has been widely investigated in previous research. 
For $\alpha=0$ and $R>R_c\approx0.34N$, 
a uniformly synchronized state is the only stable state~\cite{Wiley2006}.
For $R<R_c$, the traveling wave solution 
$$
\psi(x,t)=\Omega t+Q x, \quad Q=\frac{2\pi q}{N} %\quad (q: {\rm integer})
$$ 
where $q$ is an integer with $|q| \le \frac{N-1}{2}$,
becomes also stable.
This state is called the $q$-twisted state~\cite{Wiley2006}. 
The uniform and $q$-twisted states are replaced by  
the chimera states as $\alpha$ is increased and 
approaches $\frac12$~\cite{Kim2004,Wolfrum2011,Omelchenko2012}. 
%
%\NU{As Eq.(1) is 1D, it's not good to quote 2D and 3D results here. Please cite them in the first paragraph 
%of Introduction. The 3D case is studied by Abrams. There is also a 4D study.}
%
If the coupling range is sufficiently small compared to the system size, 
the multi-chimera states with many coherent and incoherent domains are obtained.
Statistical properties of the transition from a uniformly synchronized state to 
the multi-chimera states for $R \ll N$ 
are studied, and resemblance of the spatio-temporal patterns with directed percolation (DP) 
is pointed out~\cite{kawase2019,Duguet2019}.
DP is a two-state model of non-equilibrium critical phenomenon and describes 
onset of turbulence in various systems~\cite{Takeuchi2007,sano2016,lemoult2016}.
%
%\NU{Cite literature on DP including experimental studies.}
%absorbing state 
For the synchronized-chimera transition,
the coherent and incoherent sites correspond to inactive and active sites in DP, respectively.
For $\alpha=1$, a $q$-twisted state or coexistence of multiple $q$-twisted states (multi-twisted states) 
are obtained~\cite{Girnyk2012}.
As $\alpha$ is decreased from 1, 
chimera states appear from the multi-twisted state 
and the number of incoherent region grows. Three scenarios for the chimera birth 
has been reported~\cite{maistrenko2014cascades}. 
However, the previous studies of the twisted states are limited to a small system with 
several domains, and large-scale properties of the system with $R\ll N$ and a smaller $\alpha$ 
are not studied yet.

In this paper, we study the statistical properties 
of system (\ref{eq.main}) for a repulsive coupling ($\frac12 <\alpha < 1$) and $R \ll N$, 
exploiting an analogy to 
the directed percolation with two symmetric absorbing states (DP2)~\cite{Hinrichsen2000,Hinrichsen1997}. 
In DP2, an active site $A$ is created between two different inactive sites $I_1$ and $I_2$, 
but will never be generated from inside of an inactive region. 
Therefore the inactive state is called an absorbing state.
In 1-dimension, the active sites are interpreted as diffusing particles which annihilate by reaction $2A\to \emptyset$ and is generated by an offspring production $A \to 3A$. 
In the present model,
the $q$-twisted regions with $q>0$ and $q<0$ correspond to the two absorbing regions,
and the transition region between the two $q$-twisted region corresponds to the active site in DP2. 
We find a transition from multi-chimera state to a twisted state at a critical point $\alpha_c$ as we increase 
$\alpha$ from $\frac12$, and compare the critical behavior with that of DP2.
Further increasing $\alpha$, the multi-chimera state reappears with 
non-trivial spatio-temporal patterns. 
The fraction of incoherent sites shows a non-monotonic dependence on $\alpha$,
and we classify the spatial-temporal patterns into five phases. 

\section{Linear Stability Analysis}
    
\begin{figure}[htbp]
\includegraphics[width=\linewidth]{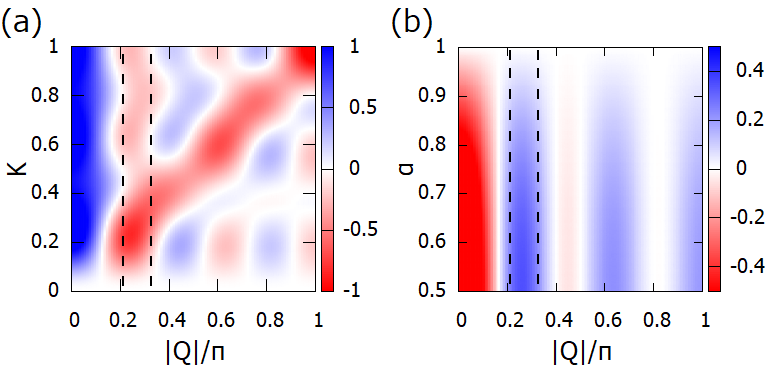}
\caption{\label{fig:1}
(a) The normalized linear growth rate ${\rm Re}\, \Lambda_K/|\cos \alpha \pi|$ 
given by Eq.~(\ref{eq:growthrate}), 
for $R=5$ and $\frac12 < \alpha \le 1$.
(b) The phase velocity of the $q$-twisted state given by Eq.~(\ref{eq:solution0}) for $R=5$.
In both plots,
the borders of the linearly stable range $0.215 < |Q|/\pi < 0.323$ are shown by dotted lines.
}
\end{figure}

In this section, we briefly recapitulate the linear stability analysis of 
the $q$-twisted states, which has been done for $\alpha = \pi$ in Ref.\cite{Girnyk2012} 
for both finite $N$ and continuum limit, and for general $\alpha$ and finite $N$ in Ref.\cite{mihara2019}.
Here we consider the case of general $\alpha$ in the continuum limit $N\to \infty$, by
adding a small perturbation to the solution for the $q$-twisted state as
\begin{equation}
\psi(x,t)=\Omega t+Q x+ \sum_K A_K e^{i K x+\Lambda_K t}.
\end{equation}
Here $A_K (\ll 1)$ and $\Lambda_K$ are the amplitude and growth rate of 
the mode with the wavenumber $K=2\pi k/N$, respectively, 
and the sum is taken over $k=1,2,\ldots, N-1$. 
Substituting this solution into the governing equation (\ref{eq.main})
and linearizing it with respect to $A_K$'s,  we get 
\begin{eqnarray}
\Omega=-\frac1{2R}\sum_{\substack{s=-R\\s\ne 0}}^{R}\sin(-Q s+\alpha\pi)
\label{eq:solution0}
\end{eqnarray}
at the zeroth order and
\begin{eqnarray}
\sum_K \Lambda_K A_K e^{iKx+\Lambda_K t}
=
-\frac{1}{2R}\sum_{\substack{s=-R\\s\ne 0}}^{R}\cos(-Q s+\alpha\pi)
\nonumber\\
\times
\sum_K A_K e^{iKx+\lambda t}(1-e^{iKs})
\label{eq:solution1}
\end{eqnarray}
at the first order, where $s=y-x$. 
From Eq.~(\ref{eq:solution1}), we extract the linear growth rate of the perturbation as
%$real part of $\Lambda_K$ as
\begin{equation}
{\rm Re} \, \Lambda_K =-\frac{1}{2R} \cos \alpha \pi
\sum_{\substack{s=-R\\s\ne 0}}^{R}\cos Q s \left(1-\cos Ks \right).
\label{eq:growthrate}
\end{equation}
Note that the sign of ${\rm Re}\, \Lambda_K$ %and therefore the linear stability 
does not depend on $\alpha$ for $\frac12 < \alpha \le 1$.
%as we see from Eq.(\ref{eq:growthrate}).
We plot the normalized growth rate ${\rm Re} \Lambda_K/(-\cos \alpha \pi)$ 
as a function of $Q$ and $K$ and for $R=5$ in Fig.\ref{fig:1}(a). 
The $q$-twisted state with $Q=2\pi q/N$ is stable if ${\rm Re}\, \Lambda_K <0$ 
for $0< K < 1$. 
The stable range is approximately given by $0.215 \pi <Q<0.323 \pi$,
which coincides with the previous result for $\alpha=\pi$~\cite{Girnyk2012}.
Note also that the stability range is extended for finite $N$ because 
the condition for stability is $\rm{Re}\, \Lambda_K<0$ for a finite number of $K$~\cite{mihara2019}. 
The phase velocity of an unperturbed $q$-twisted state for $R=5$ is plotted in Fig.\ref{fig:1}(b).
Its sign changes three times as the wavenumber $Q$ is increased, and is negative 
in the linearly stable range. The absolute value of $\Omega$ converges to zero as 
$\alpha$ approaches $1$.
%%%%%%%%%%%%%%%%%%%%%%%%%%%
%%%%%%%%%%%%%%%%%%%%%%%%%%
    
\section{Simulation Result}

\subsection{Spatio-temporal patterns}
\begin{figure}[htbp]
\includegraphics[width=8.6cm]{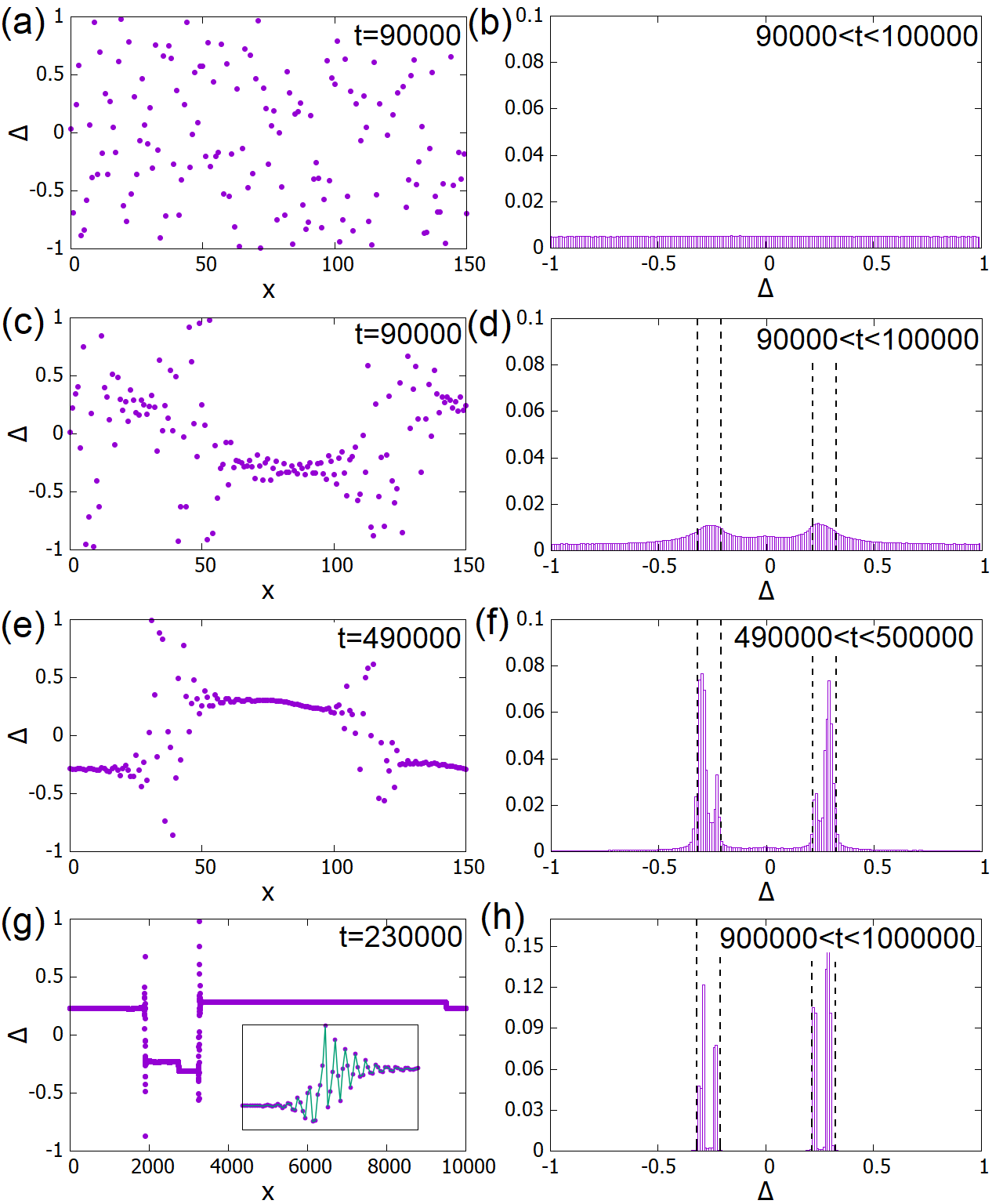}
\caption{\label{fig:2}
Spatial profiles (left) and histogram (right) of $\Delta(x)$ for (a)(b)$\alpha=0.5$, (c)(d)$\alpha=0.65$, (e)(f)$\alpha=0.70$, 
and (g)(h)$\alpha=0.73$. The plots in (a),(c),(e) show only 150 of the 10000 oscillators. 
The panel (g) shows the whole system and a magnified view of the border between 
two twisted regions (inset).
%70 sites at $x\approx3200$ 
The dotted lines in the histograms (d),(f),(h) show the borders of 
the linearly stable range $0.215 < |\Delta| < 0.323$. 
%
%\NU{Please add dotted lines to show the linearly stable range in Fig.2(d)(f)(h).}
}
\end{figure}
%%%%%%%%%%%%%%%%%
\begin{figure*}[htbp]
\includegraphics[width=17.2cm]{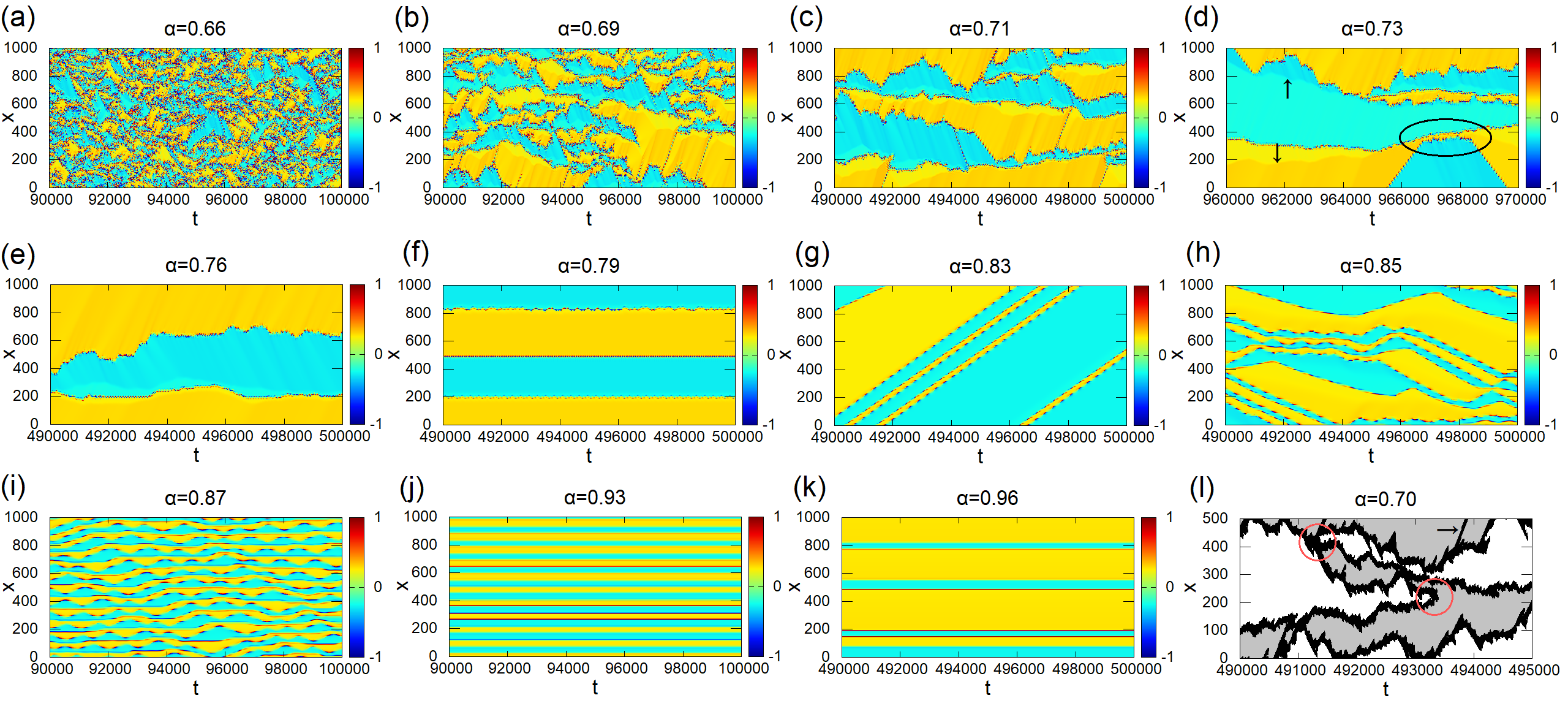}
\caption{\label{fig:3}
(a)-(k) 
Spatio-temporal patterns of the phase difference $\Delta(x,t)$ for different values of $\alpha$. 
Only 1000 out of 10000 sites and a time window of width $10000$ are shown.
Twisted regions with positive and negative phase difference 
are shown in yellow (light grey) and light blue (dark grey), 
and incoherent regions exist at the borders between them. 
In (d), the arrows show the boundaries between regions of 
slightly different $\Delta$ with the same sign.
Two incoherent regions get close and are bounced back
after a certain period of time are shown in the circle.
(l) Spatio-temporal pattern of $s(x,t)$ for $\alpha=0.70$. 
Black: $s=0$ (incoherent). White and grey: $s=\pm1$ (twisted). 
Pair-production and annihilation of incoherent sites 
are shown by red circles. Emission of a traveling wave is shown by an arrow.
%The initial condition is different from the sample in (c).
}
\end{figure*}
%%%%%%%%%%%%%%%%%
We solved the equation (\ref{eq.main}) with $\omega_0=0$ and $R=5$
for $N=10000$, using Runge-Kutta method with the time step $\Delta t=0.01$. 
The initial phases are randomly distributed in $(-\pi:\pi]$, and the periodic boundary condition is applied. 
In Fig.\ref{fig:2}, we show typical spatial profiles of the phase difference of neighbor oscillators,
\begin{equation}
\label{eq.Delta}
\Delta(x,t)=\frac{\psi(x+1,t)-\psi(x,t)}{\pi}\in(-1:1],
\end{equation}
and the normalized frequency distribution of $\Delta$.
        
When $\alpha=0.5$, the phases of oscillators are distributed uniformly, 
and so are their phase differences (Fig.\ref{fig:2}(a)(b)). 
As $\alpha$ is increased, a chimera state consisting of twisted and incoherent regions appears (Fig.\ref{fig:2}(c)). Two kind of twisted regions with positive or negative phase differences appear alternatively with an incoherent region in between.
The distribution of $\Delta$ has double peaks in the linearly stable range obtained in Sec. II.
(Fig.\ref{fig:2}(d))
As $\alpha$ is further increased, the twisted regions become larger 
and the distribution of $\Delta$ becomes sharper (Fig.\ref{fig:2}(e)-(h)).
The spatial profile of $\Delta$ in the incoherent region is oscillatory, 
and convergence to the stable value in the twisted region 
looks similar to the Gibbs phenomenon in Fourier series (Fig.\ref{fig:2}(h), inset).
    
The spatio-temporal pattern of $\Delta(x,t)$ is shown in Fig.\ref{fig:3}(a)-(k). The positively and negatively twisted regions are separated by narrow incoherent regions,
and offspring-production and pair-annihilation of incoherent regions 
(separation and coalescence of twisted regions) are observed.
        
In order to exploit the analogy with directed percolation,
we distinguish the twisted sites and incoherent sites
by using the local standard deviation of phase difference,
\begin{equation}
\label{eq.sd}
\sigma_\Delta(x,t)=\sqrt{\frac{1}{2R}\sum_{y=x-R}^{x+R-1}[\Delta(y,t)-\overline{\Delta}]^2}.
\end{equation}
If $\sigma_\Delta(x)<\sigma_c$, the site $x$ is defined as twisted, 
and if $\sigma_\Delta(x)>\sigma_c$ the site $x$ is incoherent. 
We choose $\sigma_c=0.1$ which roughly matches 
with the width of the linearly stable range of the phase difference.
% $0.215<|Q/\pi|<0.323$.
%
We define the state variable $s(x,t)=\pm1$ 
for a site with a positive and negative phase difference, respectively, 
and $s(x,t)=0$ for an incoherent site. 
A sample of the spatio-temporal patterns of $s(x,t)$ is shown Fig.{\ref{fig:3}}(l).
Interestingly, in addition to 
pair production and annihilation of the incoherent regions,
we find branching of one incoherent region into two 
and merger of two incoherent regions into one.
In these cases, one of the incoherent region is sandwiched by 
twisted regions with $\Delta$ of the same sign.
This is a difference from DP2, in which each active site is 
always located between different inactive states.
 
The density of incoherent sites is defined by
\begin{equation}
\label{eq.rho}
\rho(t)=\frac{1}{N}\sum_{x=1}^N [1- s(x,t)^2].
\end{equation}
For each simulation, $\rho(t)$ either decays to zero or fluctuates around a constant value 
after a sufficiently long time
depending on $\alpha$ and the initial condition.
        
\subsection{Absorbing-state transition}

\begin{figure}[htbp] 
\includegraphics[width=8.6cm]{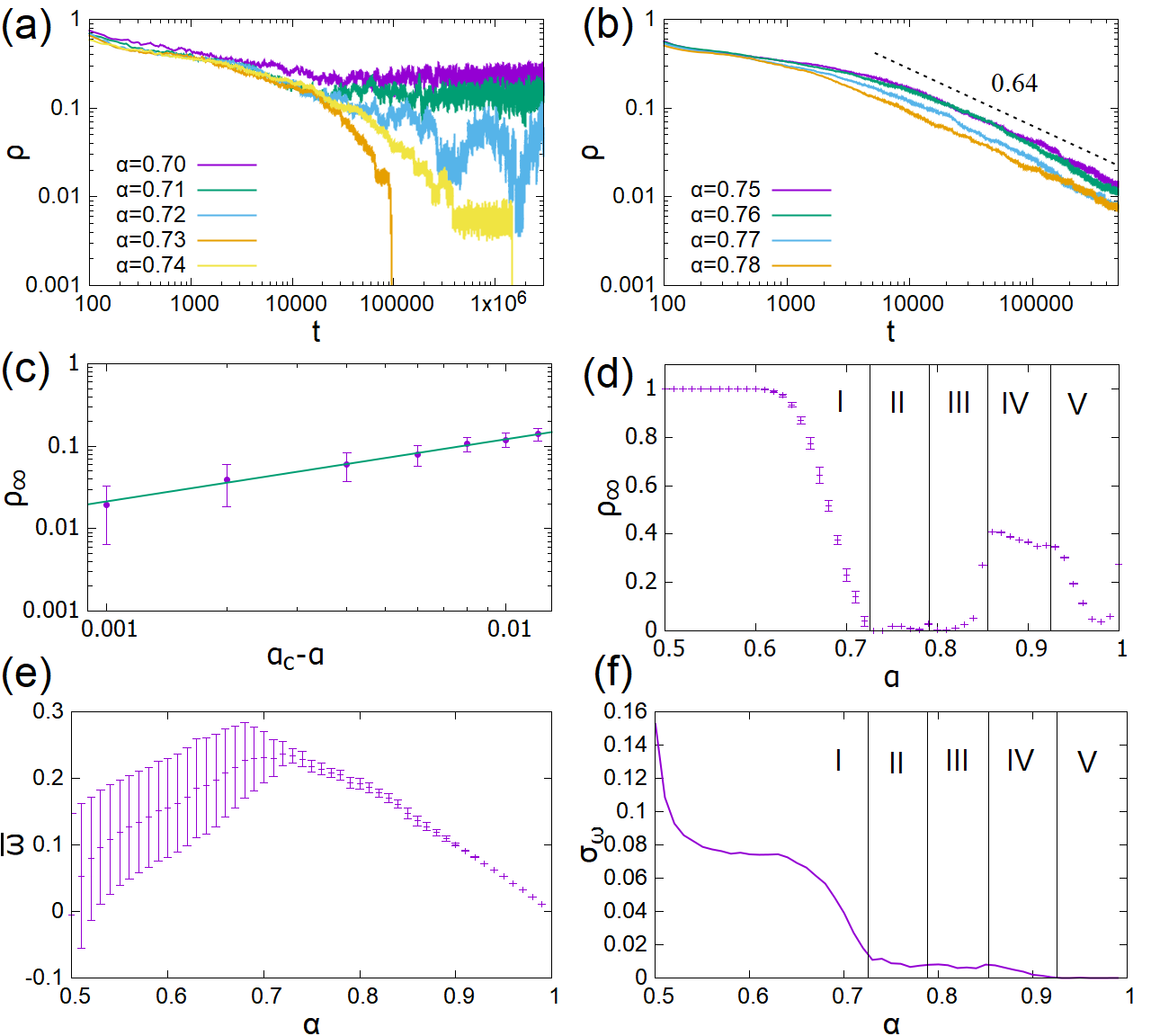}
\caption{\label{fig:4}
(a)(b) Dependence of the incoherent density $\rho$ on $t$. 
(c)(d) Dependence of $\rho_\infty$ on $\alpha$.
The error bars show the standard deviation of $\rho(t)$ for (c) $t>500000$, (d) $90000<t<100000$. 
%(a.extend 0.70 and 0.71 to t=300000 later. b.add 0.77 later. d.delete the error bar later?)
(e) The average and (f) the standard deviation of the phase velocity.
For $\alpha>0.93$, the standard deviation of phase velocity becomes zero, 
which means the transition from the multi-chimera to the multi-twisted state.
}
\end{figure}

%時空パタ－ンの説明
For $\alpha<0.79$,  the spatio-temporal patterns (Fig.\ref{fig:3}(a)-(e)) 
resemble those in DP2 that the twisted (inactive) regions grow in size and the irregularly fluctuating incoherent (active) region diminishes
as $\alpha$ is increased.
Time evolution of $\rho(t)$ for $0.70 \le \alpha \le 0.74$ is shown in Fig.\ref{fig:4}(a)(b). 
We take the time-average over the late stage and ensemble average for each $\alpha$ to 
define the asymptotic value $\rho_\infty = \rho_\infty(\alpha)$ and find that $\rho_\infty$ decreases to 0 as $\alpha$ approaches $\alpha_c=0.722$. 
Fitting by the power law $\rho_\infty \sim |\alpha - \alpha_c|^\beta$, $\beta = 0.755 \pm 0.028$ is obtained (Fig.\ref{fig:4}(c)). 
We define the states for $\alpha < \alpha_c=0.722$ as "Phase I", 
which corresponds to the active phase in DP2.
For $\alpha_c \le \alpha < 0.79$, the number of incoherent regions is either zero or a very small even number (usually 2 or 4), since the process conserves the number of incoherent regions modulo 2.
We call this state "Phase II". 
If $\rho_\infty = 0$, the whole space is occupied by one of the twisted states which satisfy the definition of "absorbing", since incoherent sites never emerge from a single twisted region.
This corresponds to the inactive phase in DP2. 
For $\alpha_c \le \alpha <0.75$, 
$\rho(t)$ decayed to zero by $t=5 \times 10^6$ in $6$ out of $7$ samples we tested. 
The survival time of the incoherent region varies.
For $0.75 \le\alpha\le 0.78$, we tested 40 samples for $t<5 \times 10^5$, 
and found that the density of incoherent sites roughly follows 
the power law $\rho(t) \sim t^{-\delta}$ with $\delta = 0.64 \pm 0.04$.
%In a later time stage where $\rho(t)<0.01$,  $\rho(t)$ changes discontinuously 
%by pair production and annihilation of the incoherent region,
%since the width $\Delta x$ of each incoherent region is roughly in the range $20 < \Delta x < 30$.
 
%%%%%%%%%%%%%%%%
%%%%%%%%%%%%%%%%
        
%%%%%%%%%%%%%%%%
\subsection{Resurgence of the multi-chimera states}

\begin{figure}[htbp]
\includegraphics[width=\linewidth]{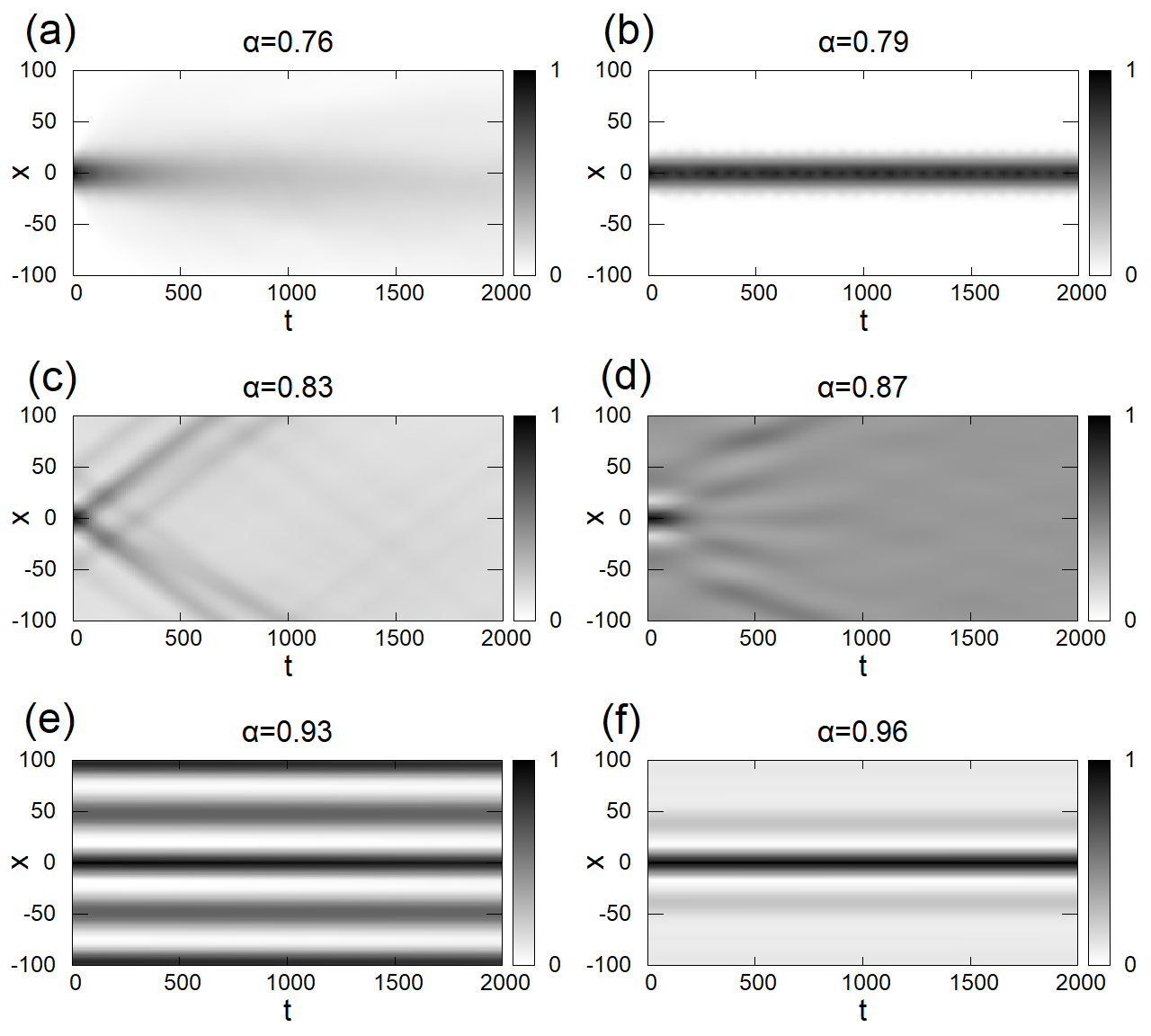}
\caption{\label{fig:5}
The space-time correlation $G(x,t)$. Gray-scale represents the strength of correlation.}
\end{figure}

We show the stationary density of the incoherent sites $\rho_\infty$
in Fig.4(d), the average of the phase velocity of all the oscillators $\overline{\omega}$ in Fig.4(e),
and its standard deviation $\sigma_\omega$ in Fig.4(f) as functions of $\alpha$.
For $\alpha\ge 0.79$, the spatio-temporal patterns of the phase difference 
changes dramatically {(Fig.\ref{fig:3}(g)-(l)}). 
We characterize them via the space-time correlation function of incoherent sites
\begin{equation}
\label{eq.corr}
G(x,t)=\langle [1-s(x',t')^2] [1-s(x'+x,t'+t)^2] \rangle_{s(x',t')=0},
\end{equation}
It gives the conditional probability that $s(x'+x, t'+t)=0$ under the condition that $s(x',t')=0$.
The space-time correlation function computed for the time window $90000 < t' < 100000$ 
is shown in Fig.~\ref{fig:5}. 
%%%%%%%%%%%%%%%%%%%%%%%%%%%%%%%%%%%%%
According to the behaviors of  $\rho_\infty(\alpha)$, $\sigma_\omega(\alpha)$ and 
the space-time correlation, we classify the dynamics of system (\ref{eq.main}) for $\alpha\ge0.79$
into phases III to V. 
    
\begin{figure}[htbp]
        \includegraphics[width=\linewidth]{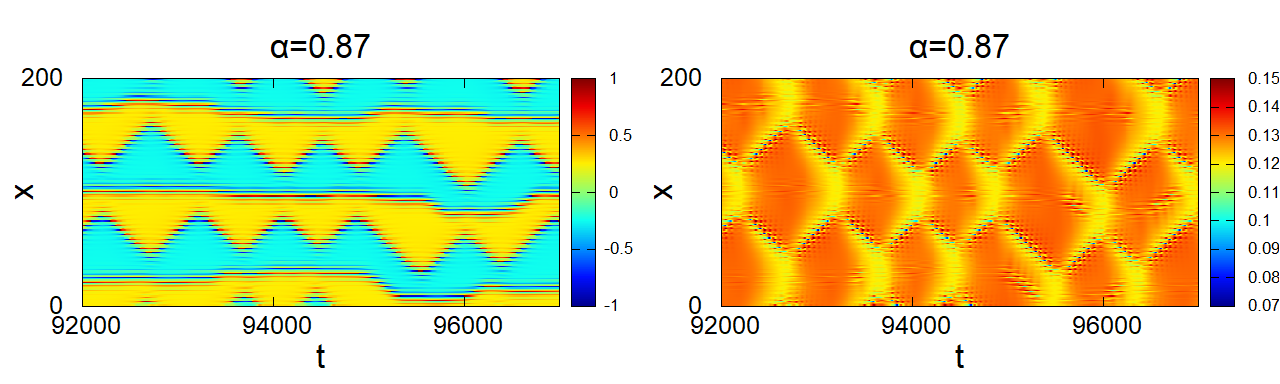}
        \caption{\label{fig:6}The spatio-temporal pattern of (a)$\Delta(x,t)$ and (b)$\dot{\psi}(x,t)$ for $\alpha=0.87$.}
        \end{figure}

Unlike the behavior in phase I and II, where the incoherent sites region diffuse from its original position (Fig.\ref{fig:5}(a)), the incoherent region maintains a straight line for $\alpha=0.79$ (Fig.\ref{fig:5}(b)). 
Phase III: for $0.80<\alpha<0.85$, the incoherent sites start to move at a constant speed at approximately 0.16 site per time unit (Fig.\ref{fig:5}(c)). The stationary density $\rho_\infty$ increases with $\alpha$ to about 0.4. 
Phase IV: for $0.85\le\alpha<0.93$, the system shows a special space-temporal pattern that straight lines and zigzag stripes coexist which is shown in (Fig.\ref{fig:6}(a)). 
%%%%%%%%%%%%%%%%%%%%%%%%%%%%%%%%%%%%%%%%%%%%
%%%%%%%%%%%%%%%%%%%%%%%%%%%%%%%%%%%%%%%%%%
 We found that the moving direction of the incoherent region is related to the phase velocity $\dot{\psi}(x,t)$ of oscillators on each side. The zigzag-moving incoherent region always moves toward the side with smaller $\dot{\psi}(x,t)$, while the $\dot{\psi}(x,t)$ on each side of the straight line has the same value (Fig.\ref{fig:6}(b)). 
For $\alpha=0.91,0.92$, the zigzag stripes begin to disappear which can been seen as a transition to the next phase.
%The density $\rho_\infty$ shows a plateau in Phase IV.
Phase V: for $0.93\le\alpha<1.00$, 
the whole system including the incoherent region between twisted regions 
evolves with the same phase velocity (Fig.\ref{fig:4}(f)). 
The incoherent regions become straight in the spatio-temporal map
while their intervals becomes larger and lose its equality as $\alpha$ is increased (Fig.\ref{fig:5}(e)(f)). 
%    We tested the case for the coupling range $R=10$ and the length of interval is doubled, which indicates that the length of interval is proportional to the coupling range $R$.

\section{Conclusion and Discussion}
%5の相に分けた1,2相は似ているが、定量的に違う
In summary, 
the spatio-temporal patterns of the non-locally coupled oscillators 
are classified into five phases depending on  $\alpha$. 
The transition from the phase I to II is qualitatively similar to the absorbing transition in DP2,
in (i) pair-production and annihilation of incoherent sites 
and (ii) vanishing of the incoherent fraction $\rho_\infty$ at the critical point.
However, some difference from DP2 are also observed. 
On one hand, every time the irregularly moving incoherent region changes its direction, a traveling wave is emitted. On the other hand, for $0.75\le\alpha\le0.78$, the decay function of $\rho(t)$ shows that the irregularly moving of  incoherent regions can not be simply interpreted as random walk, and for $\alpha<0.75$, they move in a even more complicated way which consist of both irregularly moving part and straightly moving part.
    
A fundamental difference between these two systems is that the DP2 model has only 
three possible states per site: one active state $A$ and two inactive states $I_1,I_2$, whereas the wave number $q$ in the q-twisted state can take an arbitrary integer within the stable range. 
As a result, one twisted region always contain more than one slightly different $\Delta$ (See Fig.\ref{fig:2}g, $x\approx2800,9500$ ,and arrows in Fig.\ref{fig:3}e).
The boundary of two slightly different $\Delta$ absorbs the traveling waves and prevents the annihilation of incoherent regions by repelling them back (circle in Fig.\ref{fig:3}e). 
Moreover, the critical exponent $\beta=0.755\pm0.028$ is not match to that in DP2 which is $\beta=0.90$. These evidence show that it is hard to say they belong to the same class \cite{Hinrichsen1997}.
  
For larger values of $\alpha$, we find non-monotonic change of the incoherent fraction
and the spatio-temporal patterns are classified into the phase III to V.
The non-monotonic dependence as well as the zigzag pattern in the phase IV are 
highly non-trivial and the underlying mechanism needs clarification in future work.
    
\begin{acknowledgments}
We acknowledge financial support by KAKENHI Grant Number JP21K03396 to N.U.
\end{acknowledgments}
    
%\nocite{*}

%\bibliography{reference}% Produces the bibliography via BibTeX.
\providecommand{\noopsort}[1]{}\providecommand{\singleletter}[1]{#1}%
\end{document}